\documentclass[a4paper,11pt]{article}
\usepackage{pos}

\usepackage{amsmath,multirow,bm,longtable,booktabs}
\usepackage{graphics,hyperref,subfigure,siunitx,makecell}
\usepackage{amsfonts,graphicx}

\title{Doubly charmed tetraquark: isospin channels and diquark-antidiquark interpolators}

\newcommand{\ba}{\begin{eqnarray}}
\newcommand{\ea}{\end{eqnarray}}

\author*[a]{Emmanuel Ortiz-Pacheco}
\author[b]{Sara Collins}
\author[a,c]{Luka Leskovec}
\author[d]{M. Padmanath}
\author[a,c]{Sasa Prelovsek}

\affiliation[a]{Jozef Stefan Institute,
 Jamova 39, 1000 Ljubljana, Slovenia}

\affiliation[b]{Institut für Theoretische Physik, Universität Regensburg, 93040 Regensburg, Germany}

\affiliation[c]{Faculty of Mathematics and Physics, University of Ljubljana, 1000 Ljubljana, Slovenia}

\affiliation[d]{The Institute of Mathematical Sciences, a CI of Homi Bhabha National Institute, Chennai, 600113, India}

\emailAdd{emmanuel.ortiz@ijs.si}
\emailAdd{sara.collins@ur.de}
\emailAdd{luka.leskovec@ijs.si}
\emailAdd{padmanath@imsc.res.in}
\emailAdd{sasa.prelovsek@ijs.si}

\abstract{We perform a lattice simulation to investigate the doubly charmed tetraquark $T^+_{cc}$ observed by the LHCb collaboration, slightly below the $D^{*+}D^0$ threshold, with flavor content $cc\bar{u}\bar{d}$ and isospin-$0$. Two-meson interpolators are implemented to explore the isospin quantum numbers $I=0$ and $I=1$. 
We observe attraction near the $DD^*$ threshold for $I=0$ and repulsion for $I=1$.
Moreover, we also include diquark-antidiquark interpolators to study their effect on the energy spectrum.
There is no significant shift in the ground state energy when adding diquark-antidiquark interpolators to the interpolator basis
when the heavy quark mass is close to the physical charm quark mass.
However, we observe a non-negligible shift in the
second energy level. This effect has to be
taken into account to extract the scattering amplitude of the $T^+_{cc}$.
Finally, with a higher mass (close to the bottom quark), the ground state is shifted down significantly. The simulation is performed on $N_f=2+1$ CLS ensembles with $m_\pi \simeq 280$ MeV. 
}

\FullConference{The 40th International Symposium on Lattice Field Theory (Lattice 2023)\\
July 31st - August 4th, 2023\\
Fermi National Accelerator Laboratory\\}

\begin{document}
\maketitle

\section{Introduction}

In recent years, the LHCb Collaboration has discovered numerous exotic hadrons, of which most lack a sound theoretical interpretation, and only a few of them have been studied using lattice simulations.
Among these exotic states, one, in particular, stands out: the doubly charmed tetraquark $T^+_{cc}(3875)$.	
This state was found 
only $0.36(4)$ MeV below $D^{*+}D^0$
threshold. Discovered through the invariant mass analysis of the $D^0D^0\pi^+$ decay, it exhibits a flavor content of 
$cc\Bar{u}\bar{d}$ \cite{obstcc,studytcc}. 
It was predicted by phenomenological studies to be a state with $J^P=1^+$ \cite{sym15071298}. 

The study of $T^+_{cc}$ on the lattice is related to other exotic doubly heavy tetraquarks, for example  $T_{bb}$ \cite{PhysRevD.100.014503}, and $T_{bc}$ \cite{Padmanath:2023rdu}. 
So far the lower lying energy spectrum of the doubly charmed 
tetraquark \cite{PhysRevD.99.034507, hadspec} and the $DD^*$ scattering amplitude have been 
computed \cite{PhysRevLett.129.032002,PhysRevLett.131.161901,CHEN2022137391}. For a tetraquark state it is natural to consider 
both meson-meson and diquark-antidiquark interpolators in the operator 
basis when determining the finite volume spectrum.  In particular, the 
latter are important as the heavy quark mass is increased towards the 
bottom quark mass (given that $T_{bb}$ is likely a diquark-antidiquark 
state). In this work, we compute the spectrum and energy shifts with 
respect to the $DD^*$ threshold for a heavy quark mass close to the 
charm quark mass and also in the region of the bottom quark mass. 
Results for the scattering amplitude will be presented in a future 
publication.


\section{Lattice setup and determination of eigenenergies}

 The simulation is performed on two CLS ensembles of gauge field configurations with $N_f=2+1$ dynamical quarks with $m_\pi=280(3)$ MeV and $a = 0.08636(98)(40)$ fm  \cite{bruno2015,PhysRevD.94.074501}. The ensembles differ in their spatial volumes $N_L^3=24^3$ and $N_L^3=32^3$ and they contain 255 and 492 configurations, respectively. 
 The present work investigates two values of the heavy quark masses $m_Q$ for the system  $QQ\Bar{u}\Bar{d}$, 
 \begin{itemize}
 \item $m_c$: lying slightly above the physical value of the charm quark mass, where  the masses of $D$ and $D^*$ mesons are $m_D\simeq 1927(1)$ MeV and $m_{D^*}\simeq2049(2)$ MeV, respectively.
     \item $m_{``b"}$: lying somewhat below the $b$-quark mass. This quark mass is relevant for $T_{bb}$, with  meson masses $m_{``B"}\simeq4042(4)$ MeV and $m_{``B^*"}\simeq4075(4)$ MeV.  
 \end{itemize}

In order to calculate the spectrum in the finite-volume, we calculate the 2-point correlation functions 
\ba
C_{ij}(t)=\langle 0| \mathcal{O}_i(t_{src} + t)\mathcal{O}_j^\dagger(t_{src})|0 \rangle, \label{correl}
\ea
where $\mathcal{O}_j^\dagger$ is the operator that creates states with the appropriate quantum numbers and $\mathcal{O}_i$ is the operator that annihilates the states.  Moreover, 
one can express this correlation matrix in terms of the spectral decomposition of energy levels in a finite volume, 
$C_{ij}(t)= \sum_n  Z_i^n Z_j^{*n} e^{-E_n^{lat} t},$
where $Z_i^n=\langle 0|O_i|n\rangle$ are the overlap factors. 
Using the variational method we solve the generalized eigenvalue problem (GEVP), $C_{ij}(t)u_j^{(n)}(t)=\lambda^{(n)}(t, t_0)C_{ij}(t_0)u_j^{(n)}(t)$, to determine the eigenenergies. The energies $E_n^{lat}$ are extracted from single exponential fits to the eigenvalue  correlators $\lambda^n(t)\propto e^{-E_n^{lat}}$ 
with $t_0/a=4$. 
In the following, we will present the effective energies from the diagonal correlators in Eq. (\ref{correl}), $\mathrm{E_{eff}}(t)=\mathrm{ln}\left|C_{ij}(t)/C_{ij}(t+1)\right|$, and from the eigenvalue correlators. In addition, the non-interacting $DD^*$  thresholds will be indicated. In the noninteracting limit the $DD^*$ system has discrete energies on a periodic lattice of size $L=N_L a$ given by the sum of the single-meson energies determined using the continuum dispersion relation $E^{\mbox{con}}_{H(\vec{p})}=(m_H^2+\vec{p}^2)^{1/2}$,  
\ba
E^{\mbox{ni}}=E_D(\vec{p})+E_D^*(-\vec{p}), \hspace{1cm} \vec{p}=\vec{n}\frac{2\pi}{L}, \hspace{0.5cm} \vec{n} \in N_L^3.  \label{nie}
\ea
\section{Comparing $cc\bar{u}\bar{d}$ channels with isospin $I=0$ and $I=1$}

There are two possible ways to construct color singlet tetraquark interpolators: as a system of meson-meson states and as a diquark-antidiquark configuration \cite{Close:1979bt}. We first discuss the basis of meson-meson interpolators and the spectrum of states obtained from the GEVP. The diquark-antidiquark interpolators and the impact on the spectrum when they are included in the basis is discussed in the next section.
 
 \subsection{ Meson-Meson interpolators}
 
In the meson system $(q\bar{q})$ the colour triplet quark couples with the colour antitriplet  antiquark to give the  $\mathbf{3_c}\otimes \mathbf{\bar{3}_c}=\mathbf{1_c}\oplus\mathbf{8_c}$ representations.
When considering tetraquarks as a meson-meson system $(q\bar{q})(q\bar{q})$, the only two possible combinations of two-meson $SU(3)_c$ representations that render a total color singlet state are the $(\mathbf{1_c}\otimes\mathbf{1_c})_\mathbf{1_c}$ and $(\mathbf{8_c}\otimes\mathbf{8_c})_\mathbf{1_c}$ representations.
The last representation is omitted \cite{sym15071298}, and we  
utilized the meson-meson color singlet 
$(\bar u c)_{1_c} (\bar d c)_{1_c}$.

We work in the isospin limit $m_u=m_d$, where we have either the isospin antisymmetric singlet $cc(\bar u\bar d-\bar d\bar u)/\sqrt{2}$, or symmetric triplet $cc(\bar u\bar d+\bar d\bar u)/\sqrt{2}$ flavor states under the interchange of the two light quarks.
The spin and orbital angular momentum from both the pseudoscalar $D$ meson and vector $D^*$ meson are coupled together to give a $DD^*$ system with the total angular momentum and parity $J^P = 1^+$. 

Here we have employed meson-meson interpolators of the form 
\begin{equation}
O^{DD^*}_{I}(\vec p, -\vec p)=[D(\vec{p})D^*(-\vec{p})]_{I}= \sum\limits_{\vec x_1} \bar u^a_{\alpha}(\Gamma_{1})_{\alpha\beta}e^{i\vec p_1 \vec x_1}c^a_{\beta}~\hspace{0.0cm}  \sum\limits_{ \vec x_2} \bar  d_{\delta}^b (\Gamma_{2})_{\delta\sigma}e^{i\vec p_2 \vec x_2} c_{\sigma}^b \mp \{ u\leftrightarrow d\}, 
\end{equation}
where the minus and plus sign stands for isospin $I=0$, and isospin $I=1$, respectively. The Dirac (Greek) and color (Roman) indices are explicitly written. Each meson is projected to a definite momentum $\vec{p}$, and $-\vec{p}$, such that the total momentum $\vec{P}=\vec{0}$.
We consider both $\Gamma_1=\gamma_5, \gamma_5 \gamma_t$ and $\Gamma_2=\gamma_j, \gamma_j \gamma_t$. Moreover, we utilized a sufficiently diverse basis of $DD^*$ interpolators as in Ref. \cite{PhysRevLett.129.032002}, such that they 
provide enough variety to extract the multiple energy levels reliably.
In terms of internal meson momenta, we employ interpolating operators of type $D(0)D^*(0)$ and $D(1)D^*(-1)$\footnote{For the sake of comparison between $I=0$ and $I=1$, the interpolator $D^*(0)D^*(0)$ was not included for $I=0$ in this analysis, since in the isospin channel $I=1$ it is not present for the studied irreps.}, in which two linearly independent combinations of momentum polarizations generate the partial waves $l=0,2$.
The quark fields are smeared using the  `Distillation' method, where 60~(90) Laplacian eigenvectors  were employed for $N_L=24$~($32$).

\begin{figure}
	\centering
	\begin{subfigure}{}
	\includegraphics[height=3.25cm,width=7.43cm]{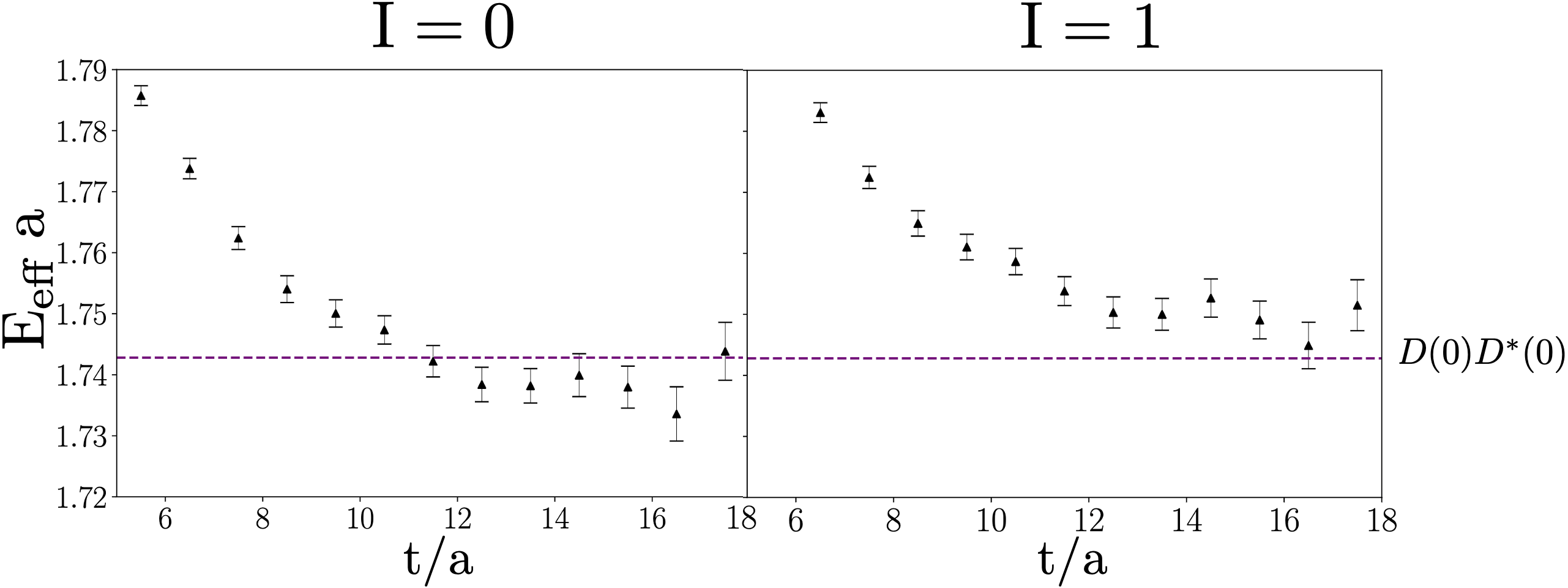}
	\end{subfigure}
	\begin{subfigure}{}
	\includegraphics[height=3.25cm,width=7.43cm]{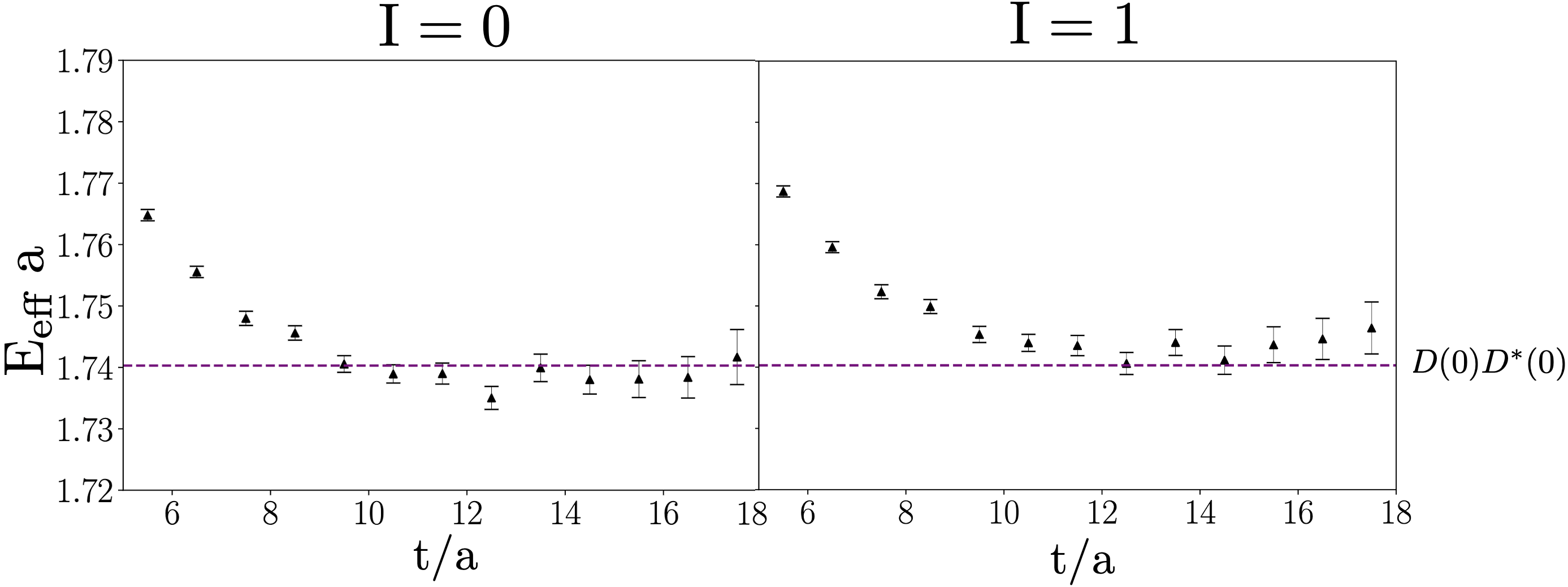}
 	\end{subfigure}
	\caption{Effective energies of the diagonal correlator constructed from the $D(0)D^*(0)$ interpolator for $J^P=1^+$ isospin $I=0$ and $I=1$ for ensemble $N_L=24$ (left) and 
 ensemble $N_L=32$ (right). Here $m_D\simeq 1927(1)$ MeV and $m_{D^*}\simeq2049(2)$ MeV.  The dashed purple lines indicate the nearest noninteracting threshold. 
        } 
\end{figure}

\begin{figure}
	\centering
	\begin{subfigure}{}
		\includegraphics[height=5.cm,width=7.35cm]{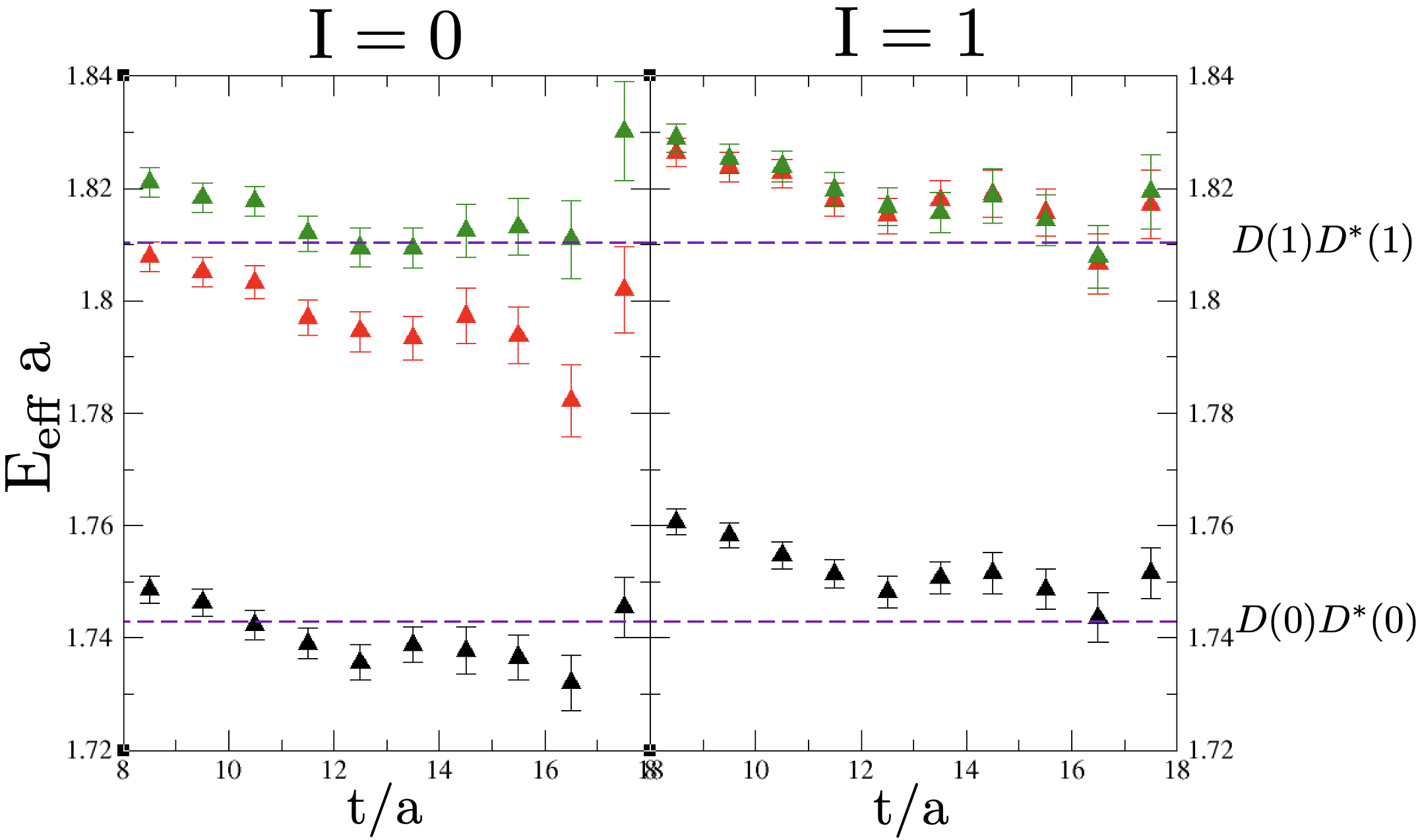}
	\end{subfigure}
	\begin{subfigure}{}
		\includegraphics[height=5.cm,width=7.35cm]{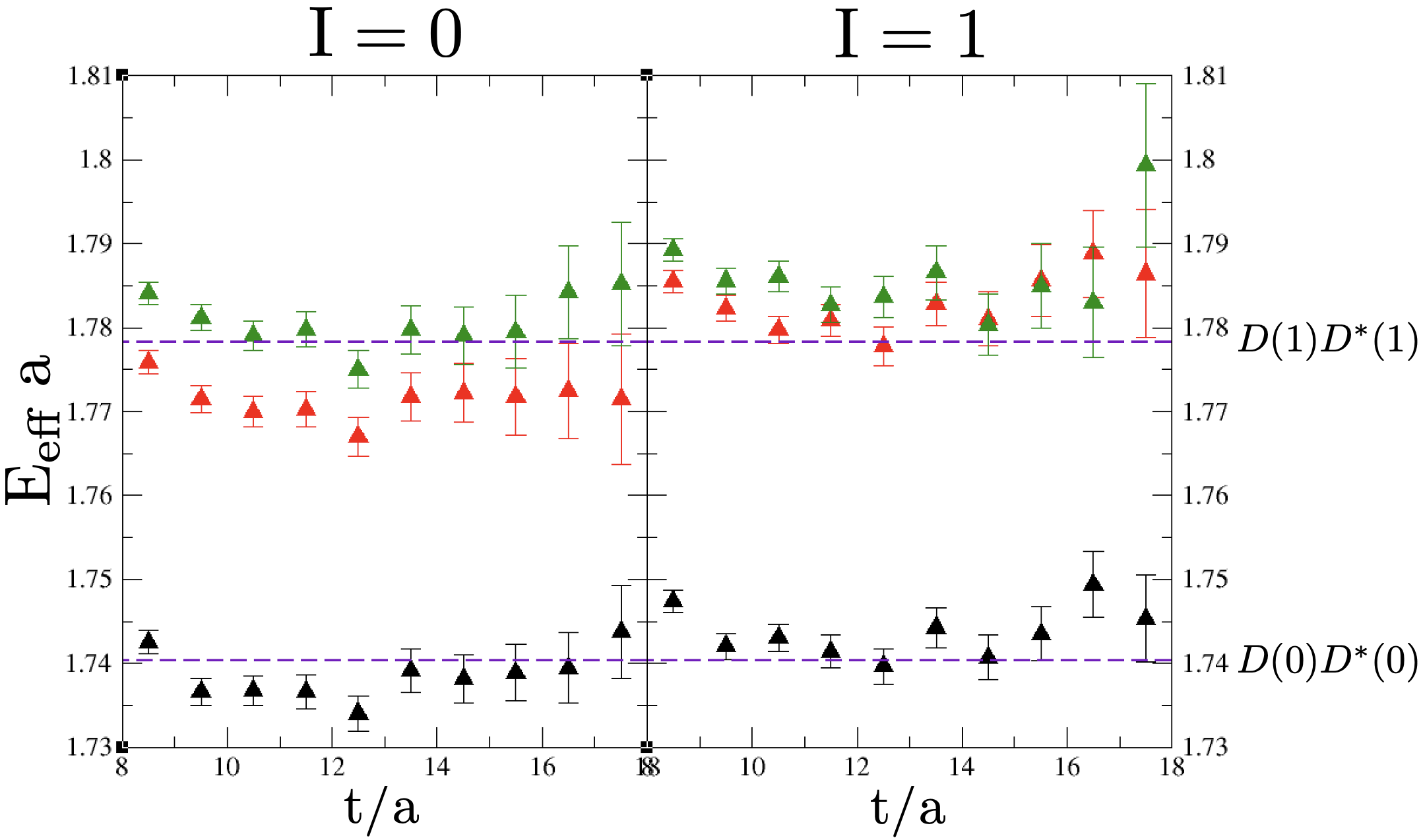}
	\end{subfigure}
\caption{Effective energies $\mathrm{E_{eff}}$ of the eigenvalues of the GEVP in the isospin channels $I=0$  and $I=1$, for ensemble $N_L=24$ (left), and for ensemble $N_L=32$ (right). The red color indicates the operator dominated by $D(1)D^*(-1)$ $s$-wave.  The noninteracting energies, from Eq. (\ref{nie}), are indicated with dashed purple lines.
}\label{gevpi} 
\end{figure}

\begin{figure}
	\centering
	\begin{subfigure}{}
		\includegraphics[height=4.cm,width=7.35cm]{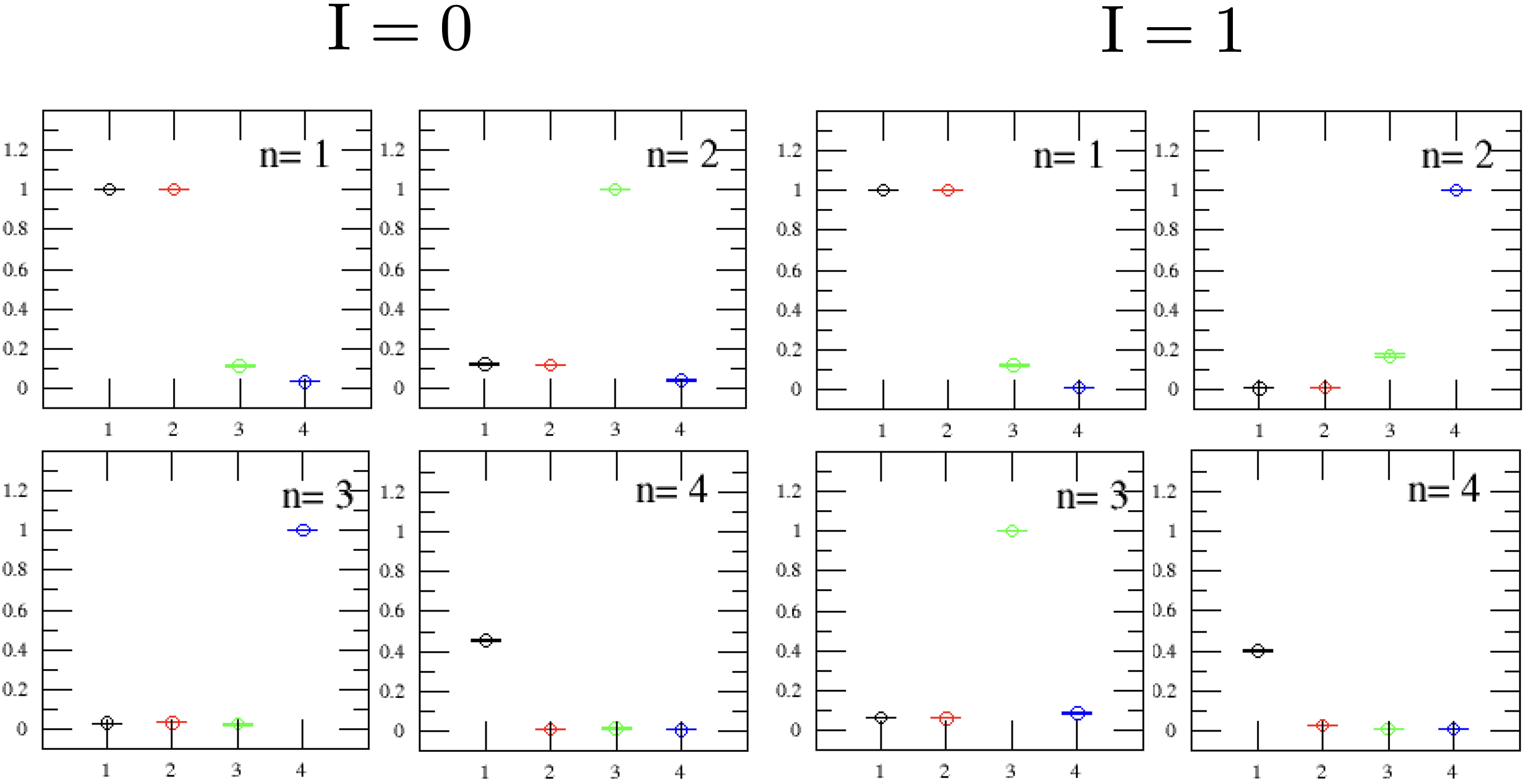}
	\end{subfigure}
	\begin{subfigure}{}
		\includegraphics[height=4.cm,width=7.35cm]{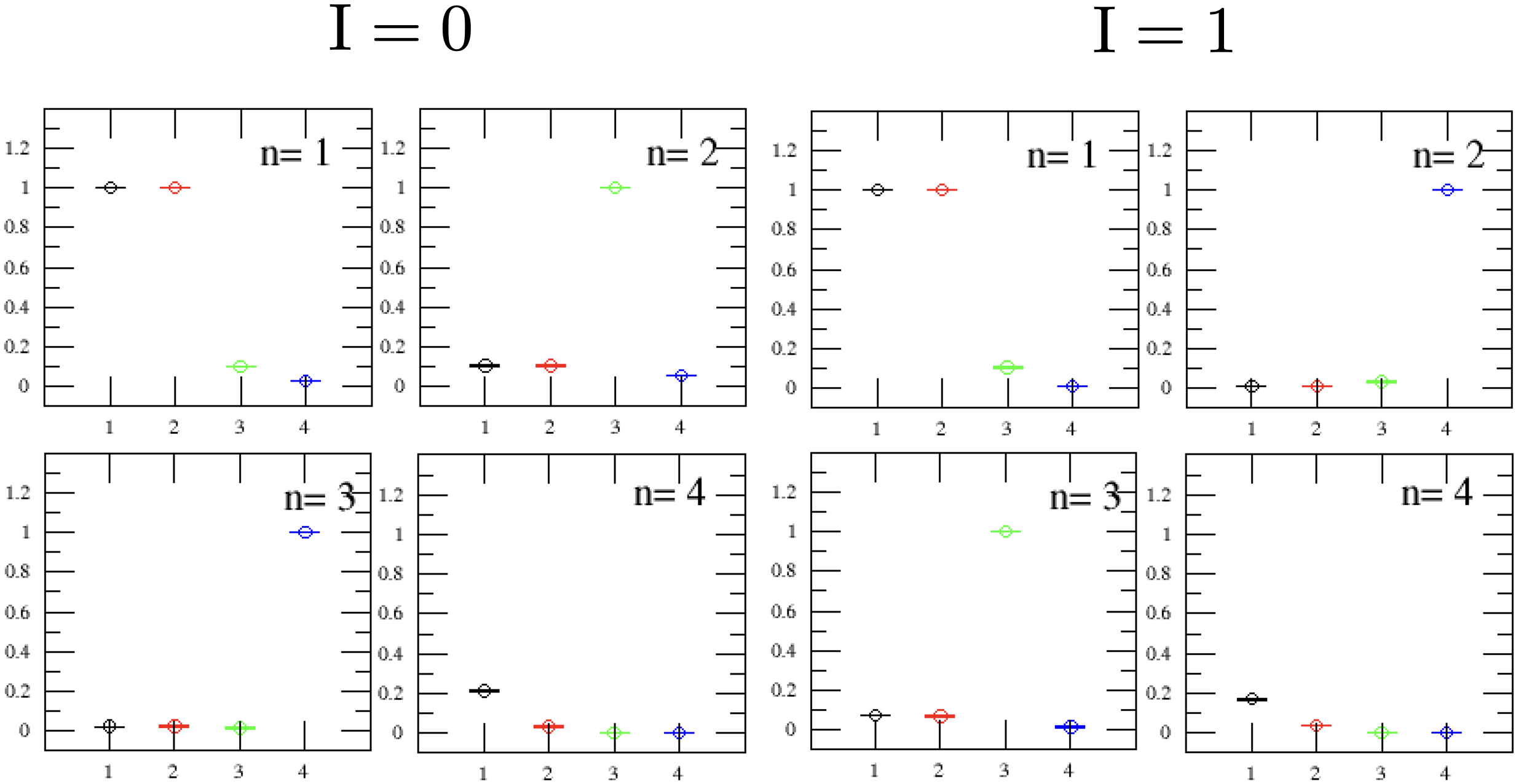}
	\end{subfigure}

	\caption{The overlap ratios $Z_j^n/\mathrm{max}_m (Z_j^m)$ of tetraquarks with $I=0$ and $I=1$ for ensemble $N_L=24$ (left), and $N_L=32$ (right). The denominator is the maximal $|Z_j^m|$ at the given operator $j$. The horizontal axis $j=1,..,4$ corresponds to operators $D(0)D^*(0)$, for both options of ($\Gamma_1=\gamma_5, \gamma_5 \gamma_t$) and ($\Gamma_2=\gamma_j, \gamma_j \gamma_t$), $D(1)D^*(-1)|_{l=0}$ and $D(1)D^*(-1)|_{l=2}$ generating  the partial waves $l=0,2$, respectively. 
 }\label{overlapsi}
\end{figure}

The effective energies of the diagonal correlators are presented in Figure \ref{correl}.  
In the plateau region the energy levels are below the noninteracting $DD^*$ threshold for $I=0$ and above for $I=1$, i.e., in the ground state we observe attraction for the $I=0$ correlators and repulsion for $I=1$. 
We observe a smaller shift for $N_L=32$ than for $N_L=24$, as expected in the larger volume.
 
 The effective eigenenergies from the GEVP analysis
 are shown in Figure \ref{gevpi} for
 the isoscalar and the isovector systems
 in both volumes. Their corresponding overlaps are presented in Figure \ref{overlapsi}.
 The ground state effective eigenenergies from GEVP are consistent with those presented in Figure \ref{correl}.
For $I=0$ the second energy level is dominated by $D(1)D(-1)$ $s$-wave, and the third level by $D(1)D(-1)$ $d$-wave.
The scattering amplitude for $I=0$ has been extracted in the previous work \cite{PhysRevLett.129.032002}, and reanalyzed by \cite{Nefediev,Baru}, while the results for $I=1$ will be presented in an upcoming manuscript.

\subsection{Wick contractions}

\begin{figure}
	\centering
	\begin{subfigure}{}
		\includegraphics[height=2cm,width=3.55cm]{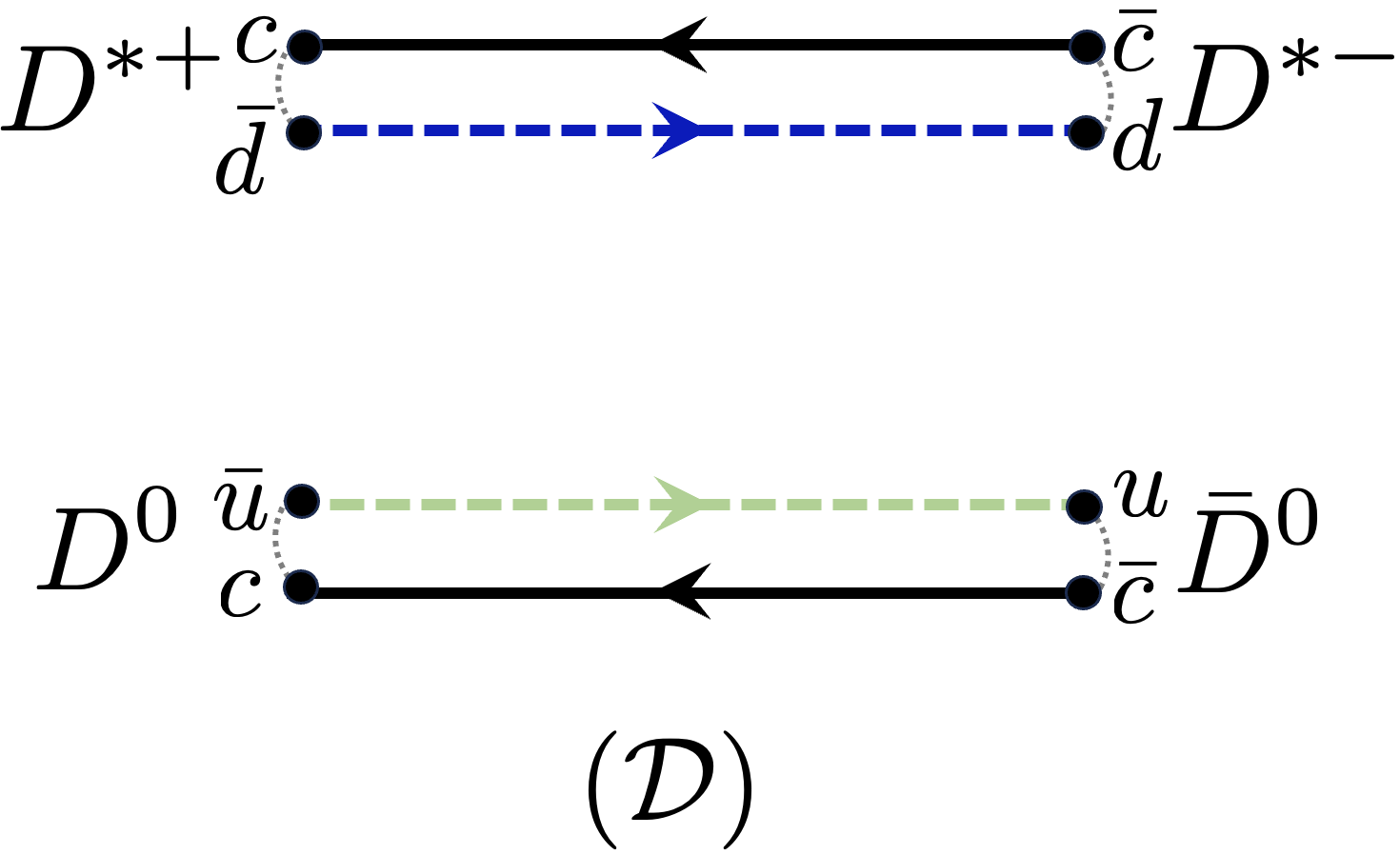}
	\end{subfigure}
	\begin{subfigure}{}
		\includegraphics[height=2.08cm,width=3.55cm]{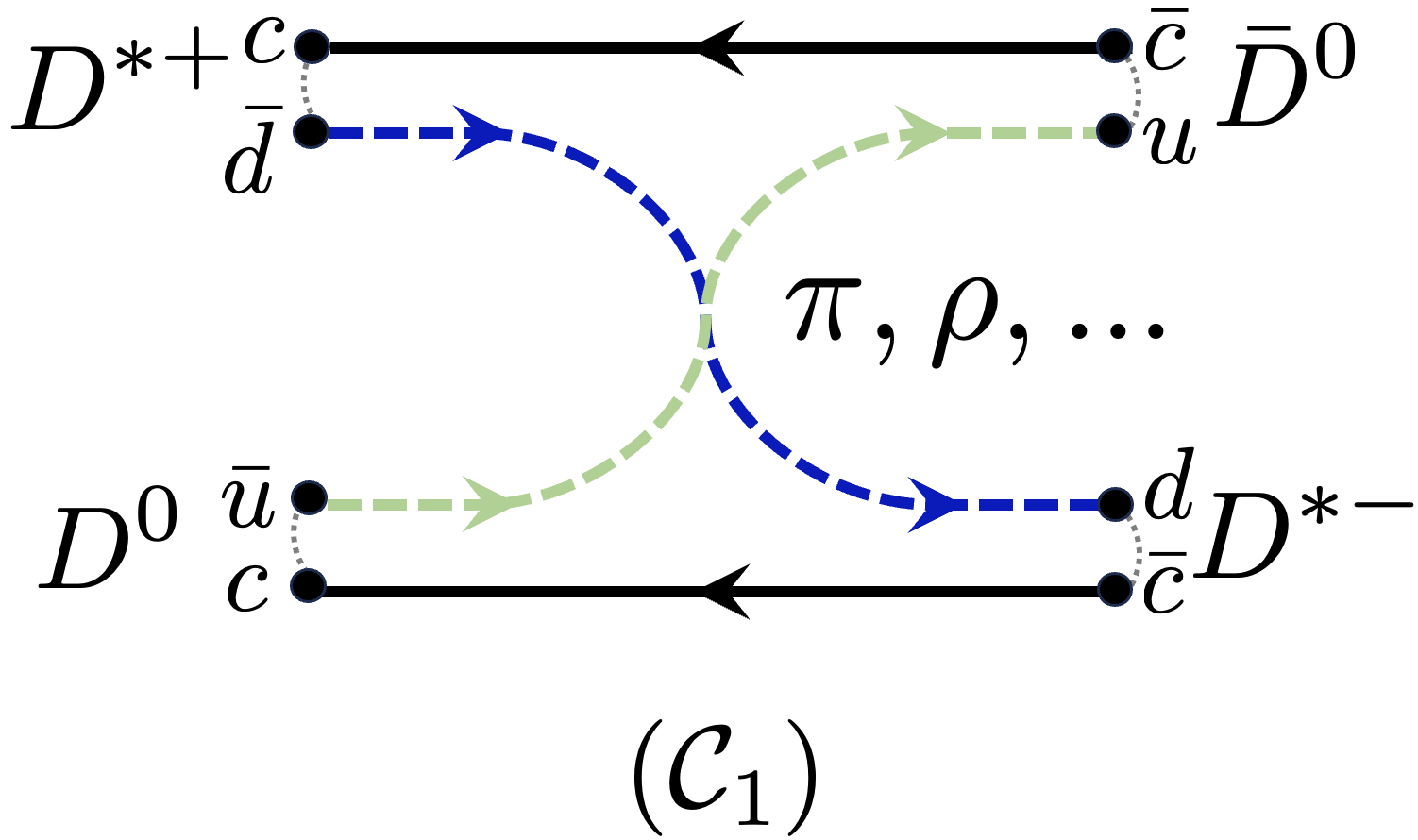}
	\end{subfigure}
 	\begin{subfigure}{}
		\includegraphics[height=2cm,width=3.55cm]{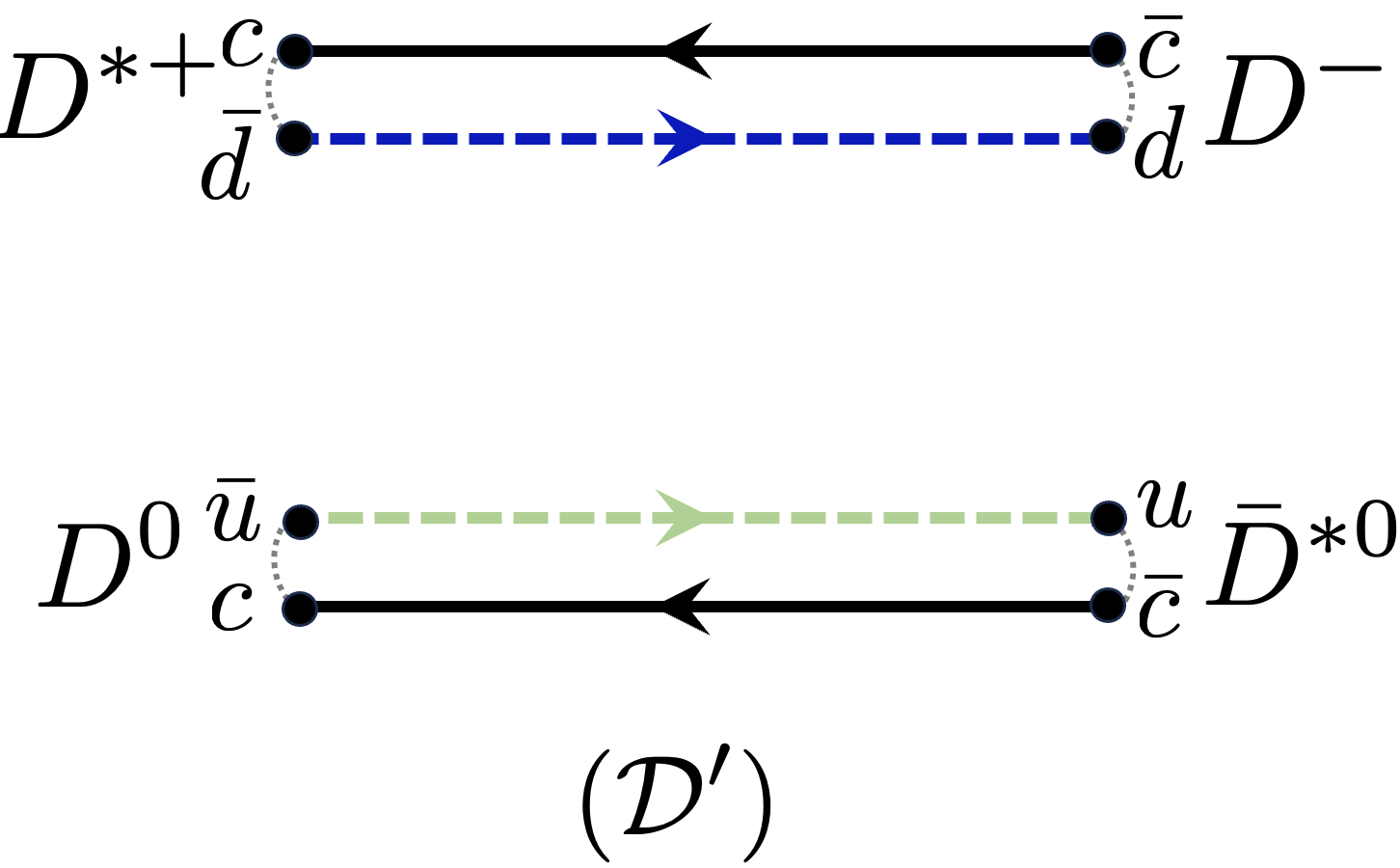}
	\end{subfigure}
	\begin{subfigure}{}
		\includegraphics[height=2.01cm,width=3.55cm]{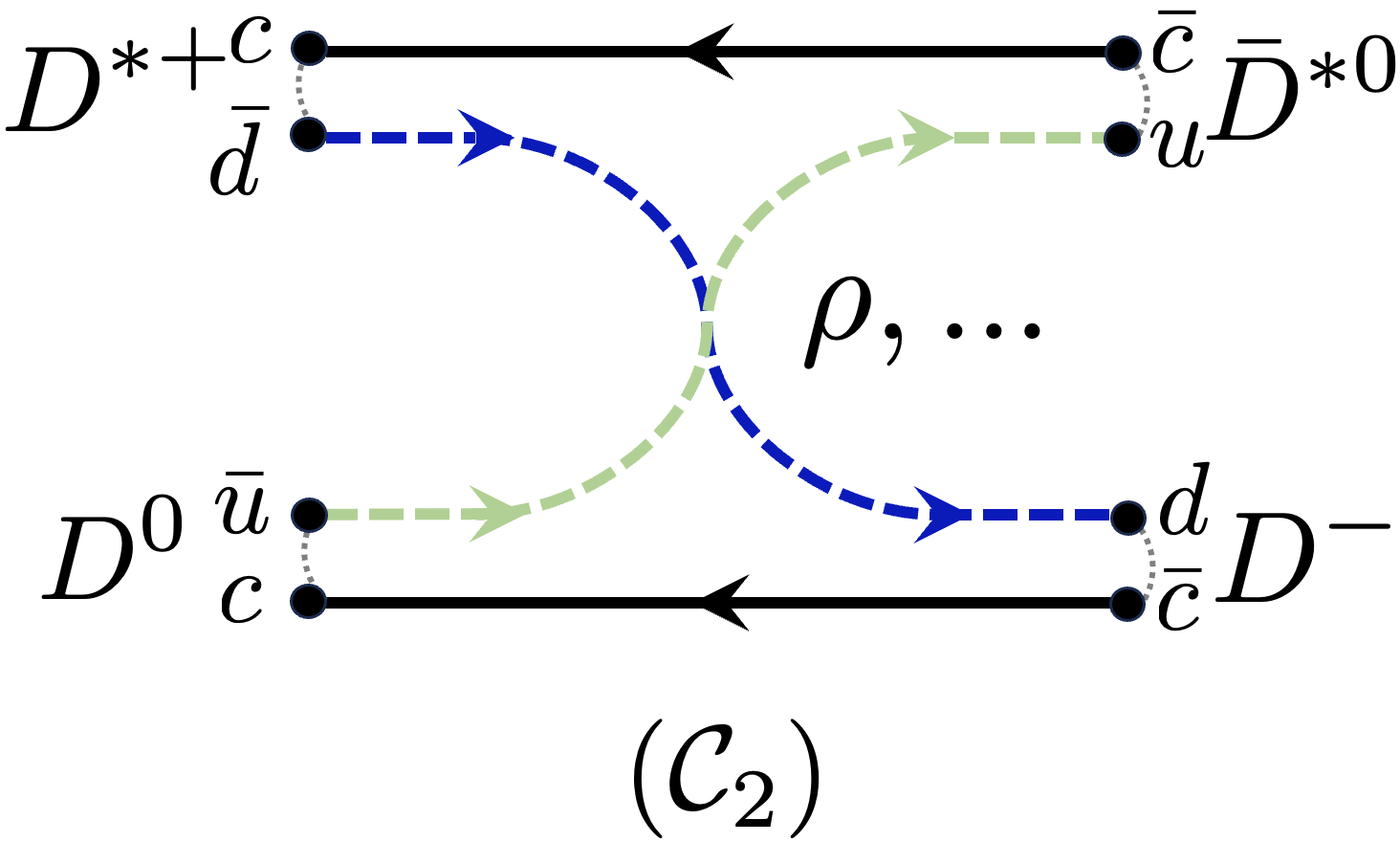}
	\end{subfigure}
	\caption{
 The four Wick (connected) contractions. The solid lines represent the charm quarks, while the dashed lines stand for the light $u$ (green), and $d$ (blue) quarks.}\label{4w}
\end{figure}

\begin{figure}
	\centering
	\begin{subfigure}{}
		\includegraphics[height=4.4cm,width=5.9cm]{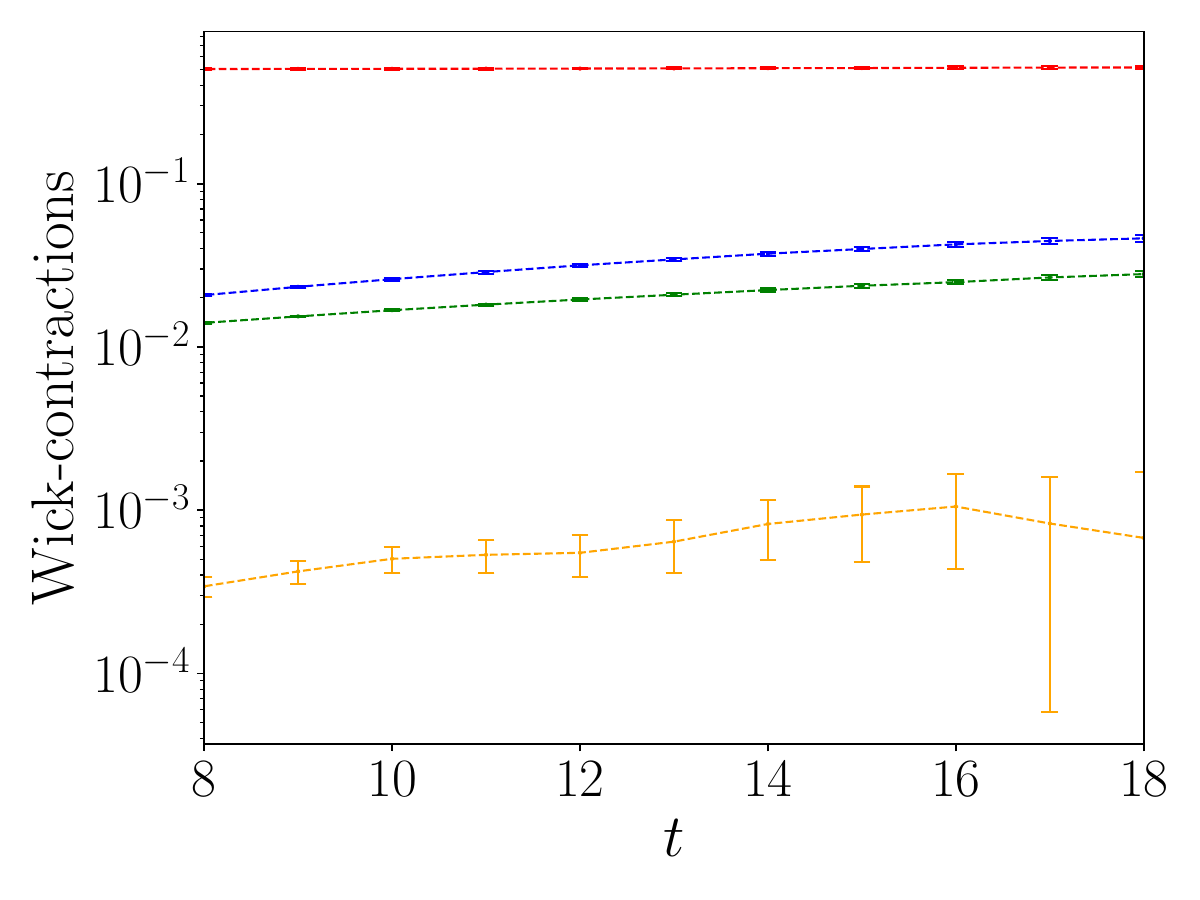}
	\end{subfigure}
	\begin{subfigure}{}
    \includegraphics[height=4.6cm,width=5.9cm]{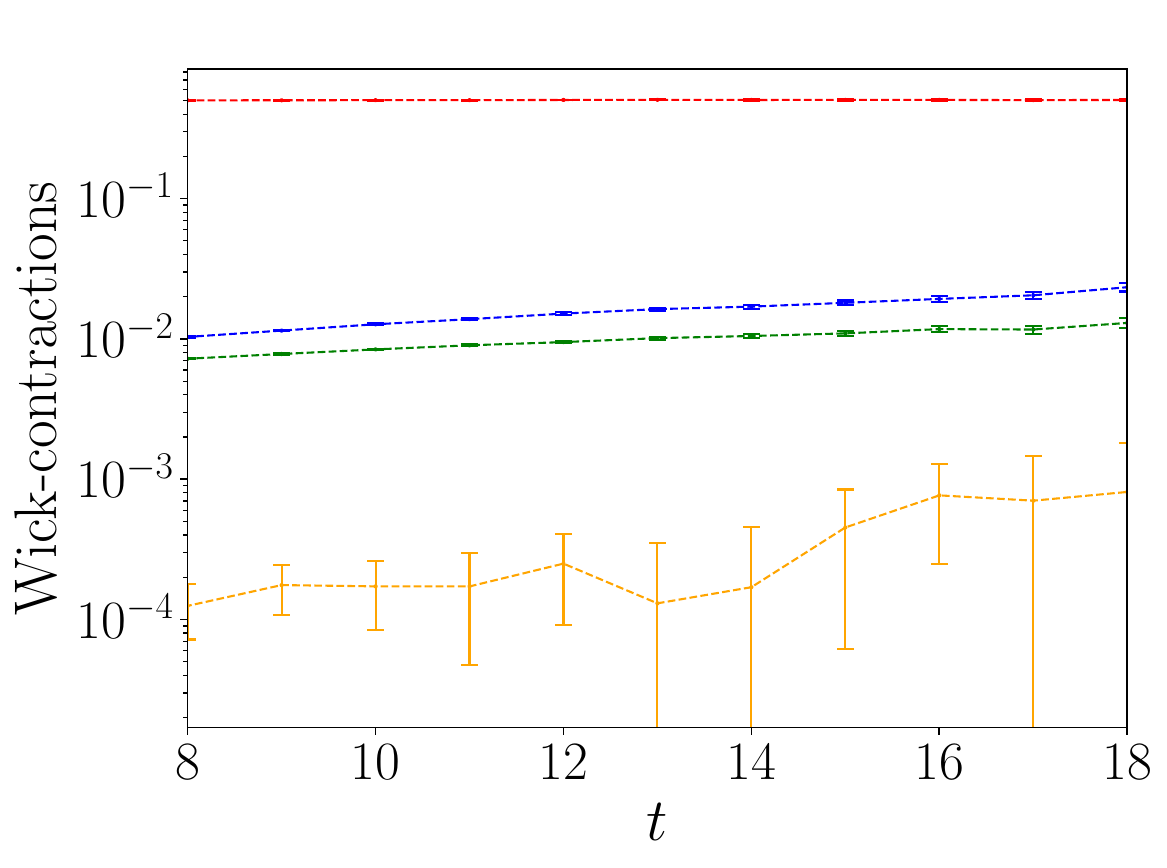}
	\end{subfigure}
     \raisebox{1.8cm}{
	\begin{subfigure}{}
    \includegraphics[height=1.2cm,width=2.35cm]{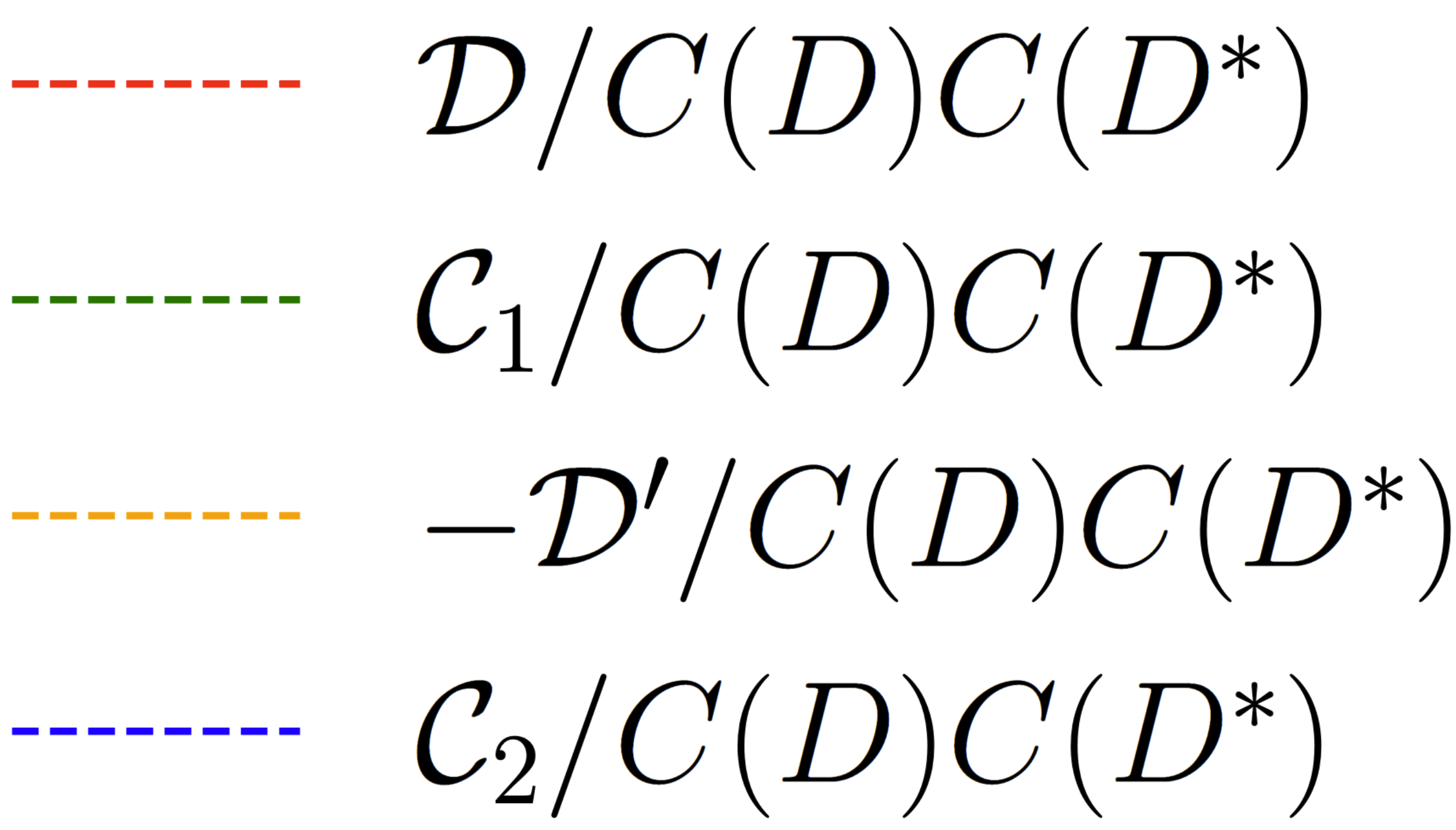}
	\end{subfigure}
    }
 \caption{Contribution of each of the four Wick contractions (displayed in Figure \ref{4w}) to the correlation function from the $J^P=1^+$ $(T_1^+ \:\hbox{irep})$ in the rest frame ($|\vec{P}|=0$) for the ensemble $N_L=24$ (left), and ensemble  $N_L=32$ (right). Each Wick contraction is normalized by the product of the single-meson $D$ and $ D^*$ correlators. }\label{wc}
\end{figure}

Similar to the lattice study performed in \cite{CHEN2022137391}, we analyze the four Wick contractions  arising from the correlators in Eq. (\ref{correl}), depicted in Figure \ref{4w}. We study each of the Wick contractions separately for the diagonal correlators built from $D(0)D^*(0)$ interpolators in the finite volume irrep $T_1^+$. This results in the following  expression for the correlators depending on the isospin quantum number (following the notation in Ref. \cite{CHEN2022137391})
\ba
C^{DD^*}_{I}(t)= \mathcal{D}-\mathcal{C}_1+(-)^{I+1}(\mathcal{D'}-\mathcal{C}_2).
\label{esr3}
\ea
Contractions $\mathcal{D}$ and $\mathcal{C}_1$ appear with the same sign, and they give the same contributions for both $I=0$ and $I=1$,  while the contribution from  $\mathcal{D'}$ and $\mathcal{C}_2$ depends on the isospin channel.  
Every contribution to the $C^{DD^*}_{I}$ correlator function from Eq. (\ref{esr3}) is computed, and the results are shown in Figure \ref{wc}. Contraction $\mathcal{D}'$ gives the smallest contribution by at least one order of magnitude compared to the rest of the contractions, while contraction $\mathcal{D}$ makes the largest contribution. The crucial Wick contraction differentiating between the isospin channels corresponds to $\mathcal{C}_2$ (Figure \ref{wc}). 

The diagrams from Figure \ref{4w} can be understood as follows: 
Both $\mathcal{D}$ and $\mathcal{D'}$ diagrams, have direct propagators connecting $D$ and $D^*$ mesons, either of the same or different types of mesons, respectively.
The $\mathcal{C}_1$ and $\mathcal{C}_2$ diagrams exhibit explicit light quark $u$ and $d$ exchanges between $D$ and $D^*$ mesons, which can be seen as pseudoscalar-, or vector-meson exchanges. Nevertheless, in diagram $\mathcal{C}_2$ the interaction $DD\pi$ is $\mathcal{P}$-parity forbidden, and  $\rho$ or heavier meson exchange may be the mediators.
Evidently, the contribution from $\mathcal{C}_2$ differentiates the interactions within the two isospin channels. This observation is consistent with that of the previous lattice study in Ref.  \cite{CHEN2022137391}. 
Currently, efforts are being made to extract the relevant scattering amplitude near the $DD^*$ threshold for $I=1$.

\section{Impact of diquark-antidiquark interpolators on $T^+_{cc}$ with $I=0$}
In a diquark $[qq]$, both quarks couple together to give the
$\mathbf{3_c}\otimes\mathbf{3_c}=\mathbf{6_c}\oplus \mathbf{\bar{3}_c}$ color representations of $SU(3)_c$, whereas the coupling of color in the antidiquark  $[\bar{q}\bar{q}]$ displays  the  $\mathbf{\bar{3}_c}\otimes\mathbf{\bar{3}_c}=\mathbf{\bar{6}_c}\oplus \mathbf{3_c}$ representations. Therefore, 
the two possibilities to couple the diquark with the antidiquark to get the color singlet tetraquark states are either the $(\mathbf{\bar{3}_c}\otimes\mathbf{3_c})_\mathbf{1_c}$ or $(\mathbf{6_c}\otimes\mathbf{\bar{6}_c})_\mathbf{1_c}$ representations. Both $\mathbf{3_c}$ and $\mathbf{\bar{3}_c}$ are antisymmetric under the permutation of the two color labels, whereas both $\mathbf{6_c}$ and $\mathbf{\bar{6}_c}$ are symmetric.

For a diquark-antidiquark tetraquark, many studies favor the triplet-antitriplet representation of the tetraquark $[cc]_{\bar 3_c} [\bar u\bar d]_{3_c}$  to be the relevant structure to provide the energy spectrum \cite{triplet}
\ba
 O^{4q}_{I=0}(\vec P)=\sum\limits_{\vec x} \epsilon_{abc}    c^b_\alpha(\vec x) (C \gamma_i)_{\alpha\beta} c^c_\beta(\vec x)  \hspace{0.2cm}  \epsilon_{ade}    \bar u^d_\delta(\vec x) (C \gamma_5)_{\delta\sigma} \bar d^e_\sigma(\vec x)  e^{i\vec P \vec x},
 \ea
where $C$ is the charge conjugation matrix and  the Roman and Greek indices represent the color and Dirac indices, respectively.
The heavy diquark $cc$ is flavor symmetric, while the light antidiquark $\bar{u}\bar{d}$ is flavor antisymmetric in $I=0$. 
Hence to satisfy Fermi symmetry, the spin has to be symmetric for the heavy diquark ($S=1$), while antisymmetric for the light antidiquark ($S=0$).
Furthermore, we also include the same set of interpolators as in the study of the meson-meson interpolators
 with the important exception that in the present section  we also consider $D^*(0)D^*(0)$ operators in $T^+_1$. 

\begin{figure}
	\centering
	\begin{subfigure}{}
		\includegraphics[height=7.5cm,width=14cm]{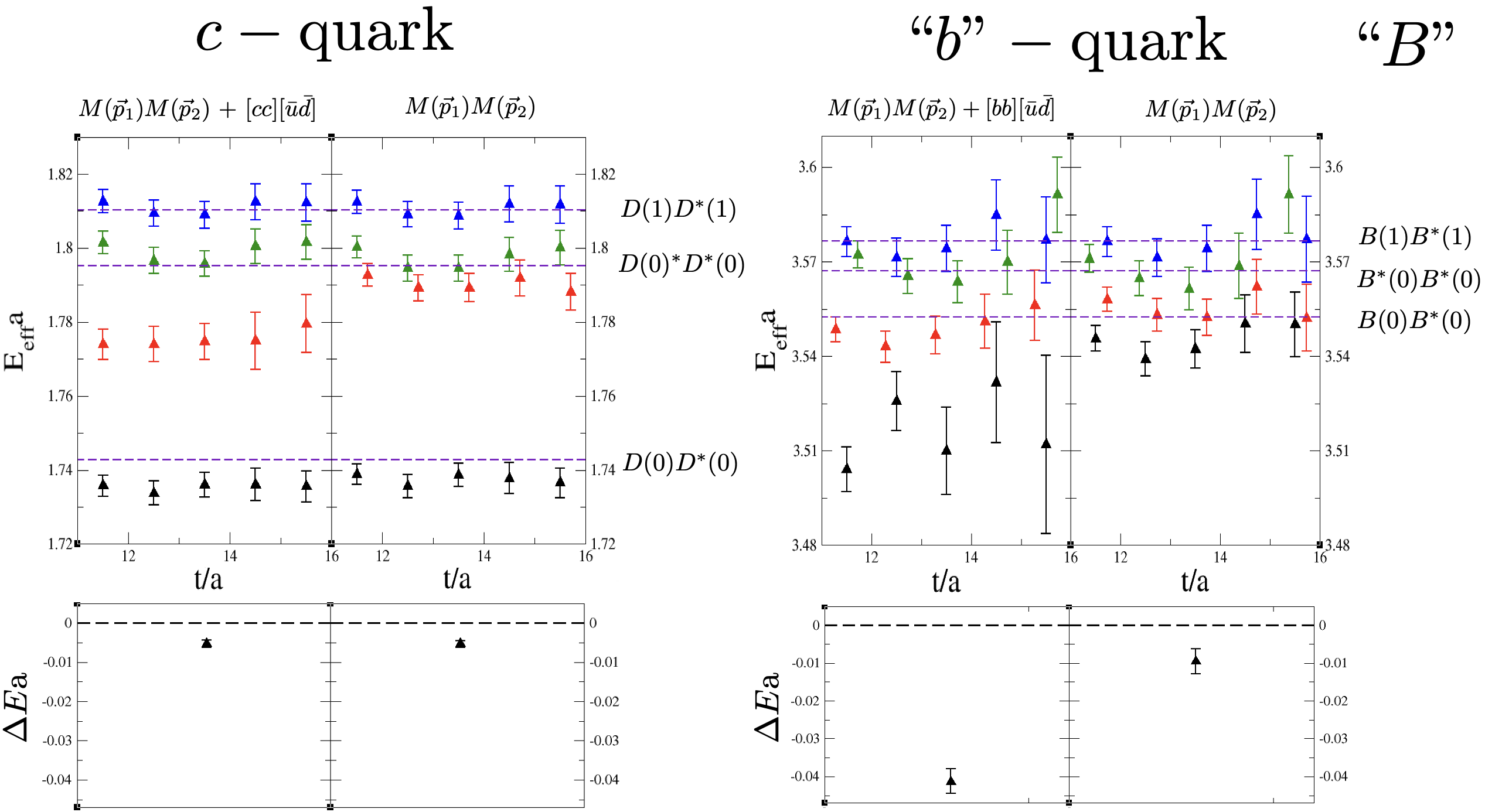}
	\end{subfigure}
 	\caption{Comparison between the effective eigenergies  for the doubly heavy tetraquark $QQ\bar{u}\bar{d}$ with $I=0$  obtained from the GEVP utilising a basis of meson-meson interpolators $M(\vec{p}_1)M(\vec{p}_2)$ and an extended basis including diquark-antidiquark operators $[QQ][\overline{u}\overline{d}]$. We study two heavy quark masses $m_Q$, one near to the physical charm quark mass (left), and the other one close to the $b$-quark mass (left). The corresponding energy shifts of the ground state are shown at the bottom.}\label{gevp4q}
\end{figure}

The effective eigenergies of the doubly charmed tetraquark obtained by utilising diquark-antidiquark interpolators in addition to meson-meson operators are presented in Figure \ref{gevp4q}. 
When we compare the ground state from meson-meson interpolators with the one including diquark-antidiquark interpolators, we do not observe a significant shift at $m_Q\simeq m_c$.  
This is reflected in the comparison of both energy shifts in  Figure \ref{gevp4q}.  The energy shifts were determined as the difference $\Delta E=E_n^{lat}-E^{lat}_{D(\vec{p_1})}-E^{lat}_{D^*(\vec{p_2})}$.   
Turning to the  $``b"$-sector, the thresholds are closer to each other.
The ground state energy level is observed to shift significantly downwards with the inclusion of diquark-antidiquark interpolators.
 This effect is also observed in the energy shifts of the ``$b"$-sector in Figure \ref{gevp4q}. This result is consistent with previous lattice studies of $T_{bb}$ with diquark-antidiquark interpolators  \cite{PhysRevD.100.014503}, where the signature of a deeply bound state is observed. 

Concerning the excited energy levels, we have observed an impact on the second energy level at $m_Q\simeq m_c$ once the additional diquark-antidiquark interpolators are employed (Figure \ref{gevp4q}). This energy shift cannot be disregarded and must be consistently considered in the extraction of the scattering amplitude near $DD^*$ threshold, relevant for $T^+_{cc}$. So far, the scattering amplitude has never been extracted including diquark-antidiquark interpolators in addition to meson meson operators.

\section{Conclusions and outlook}

In the present lattice simulations, we explore both doubly charmed tetraquark  states $I(J^P)=0(1^+)$ and $1(1^+)$ by implementing meson-meson interpolators. 
Attraction in the $I=0$ channel and repulsion in the $I=1$ channel near the $DD^*$ threshold are observed. Furthermore, we confirm the identification of the relevant Wick contraction and the corresponding diagram responsible to differentiate the isoscalar from the isovector channel. 
Our observations are consistent with that made in Ref. \cite{CHEN2022137391}. 
Currently, we are in process of extracting the scattering amplitude in the isovector channel.

When we additionally consider diquark-antidiquark interpolators with $I=0$ for a heavy quark mass close to the physical  $c$-quark mass, no significant shift in the ground state energy is found. However, regarding the remaining energy levels, we  observe a non-negligible shift in the second energy level once we include diquark-antidiquark interpolators. This effect has to be taken into account to extract consistently the scattering amplitude of the $T^+_{cc}$. 
Finally, when we tune the mass of the heavy quark to be close to the physical ``$b"$ quark mass, we find a significant energy shift downwards with respect to the $BB^*$ noninteracting threshold, once the diquark-antidiquark interpolators are used in addition to meson meson operators. This result is consistent with studies of $T_{bb}$ on the lattice. 


\section{Acknowledgements}

The authors thank the HPC RIVR consortium (www.hpc-rivr.si) and EuroHPC JU (eurohpc-ju.europa.eu) for funding this research by providing computing resources of the HPC system Vega at the Institute of Information Science (www.izum.si) and the Department of Theoretical Physics at JSI for the use of the HPC Vega system and the ATOS cluster. E.O.P, L.L. and S.P. acknowledge the projects (J1-3034) and (P1- 0035) were financially supported by the Slovenian Research Agency ARRS.



\bibliographystyle{apsrev4-1}
%

\end{document}